# Different Melting Behavior in Pentane and Heptane Monolayers on Graphite; Molecular Dynamics Simulations

**Cary L. Pint**

Department of Physics, University of Northern Iowa, Cedar Falls, IA  50614-0150

## ABSTRACT

Molecular dynamics simulations are utilized to study the melting transition in pentane ($C_5H_{12}$) and heptane ($C_7H_{16}$), physisorbed onto the basal plane of graphite at near-monolayer coverages.  Through use of the newest, optimized version of the anisotropic united-atom model (AUA4) to simulate both systems at two separate coverages, this study provides evidence that the melting transition for pentane and heptane monolayers are significantly different.  Specifically, this study proposes a very rapid transition from the solid crystalline rectangular-centered (RC) phase to a fluid phase in pentane monolayers, whereas heptane monolayers exhibit a transition that involves a more gradual loss of RC order in the solid-fluid phase transition, accompanied by uniaxial motion of sublattices and molecules within sublattices.  Through a study of the melting behavior, encompassing variations where the formation of gauche defects in the alkyl chains are eliminated, this study proposes that this melting behavior for heptane monolayers is a result of less orientational mobility of the heptane molecules in the solid RC phase, as compared to the pentane molecules.  This idea is supported through a study of a nonane monolayer, which gives the gradual melting signature that heptane monolayers also seem to indicate.  The results of this work are compared to previous experiment over pentane and heptane monolayers, and are found to be in good agreement.

# I. INTRODUCTION

The study of thin films of physisorbed molecules has been a topic that has received a significant amount of both experimental and theoretical attention in the past few decades. The importance of these types of systems is becoming more realistic as the technology to probe and study them on an increasingly small length scale evolves in concert with computing performance which permits larger and longer simulations to study that which is still not observable through even some of the most detailed experimental work.

One particular topic that has received a significant amount of attention and interest drawn toward it has been the adsorption of *n*-alkanes onto the basal plane of graphite. Since the *n*-alkanes are fundamental constituents of many products and substances that are very important for industrial and technological application, this family of molecules is at the forefront of many surface applications that range over topics that involve detergency, adhesion, lubrication, and many other similar applied fields. An understanding of the short-chained alkanes on a surface is sought due to their well-known lubrication and adhesion properties, which provides insight into better lubricants and adhesives for industrial use. Furthermore, the *n*-alkanes have been found to be fundamentally important in theoretical studies of more complex systems, such as lipid bilayers,[1-2] whose intramolecular motions occur over very long time scales which make these complex systems very arduous to simulate. Therefore, what is currently understood about this family of molecules in physisorbed systems has proven to play a significant role in the understanding of many surface systems and surface problems, and one can only imagine where a full understanding of the behavior of these *n*-alkanes on surfaces will lead in the future.

The systems involved in this work deal with primarily two short-chained *n*-alkane molecules, pentane ($C_5H_{12}$ or C5) and heptane ($C_7H_{16}$ or C7), physisorbed onto the basal plane of graphite at two previously experimentally determined and studied coverages that are near to full monolayer completion. To further study the effects of chain length, nonane ($C_9H_{20}$ or C9) monolayers are studied since their phase behavior is similar to that exhibited by pentane and heptane monolayers. Due to its attractive properties and its popularity in previous experimental work, graphite is utilized for the substrate in this study. The importance of physisorbed $n$-alkanes on graphite is emphasized by the vast experimental effort in recent years[3-10] to study the solid and liquid-state behavior in such systems. These studies have found that for even-numbered short-chained $n$-alkanes[3-5], with $n<12$, the monolayer arranges in a low-temperature herringbone (HB) phase that is commensurate with the substrate (except for butane, which is not found to exhibit a HB phase). Upon heating, these studies report that an intermediate phase, where a solid phase coexists with a liquid, forms until the monolayer undergoes a melting transition. In the case of pentane, this coexistence region is found to be quite small, and the melting temperature of the coexisting monolayer is found to be very near to the bulk melting temperature.

Unlike the short-chained even $n$-alkanes, the odd alkanes[5-10] show significantly different low-temperature solid phase behavior. Where the even alkanes form a low-temperature HB phase, the odd alkanes form a low temperature rectangular-centered (RC) phase that is at least partially commensurate with the graphite substrate. In particular, the study by Matthies[6] studies the C5 and C7 monolayers through X-ray and neutron diffraction at reported coverages of 1.01 and 0.98 monolayers respectively. This study finds the RC phase present, but proposes that there is ambiguity in the diffraction patterns regarding fits involving a slightly rotated HB phase. With aid of temperature dependent diffraction patterns, this study observes a shift in the Bragg peak diffraction data that is interpreted as the melting transition, and it is observed that this transition is very rapid and occurs between ca. 99-105K. For C7 on graphite, this study finds that the monolayer seems to undergo a more augmented coexistence region than that observed for C5, with a melting transition occurring at about $T$=170K for the 0.98 monolayer sample. In both cases, a proposed model with a slight HB rotation is proposed, and will be utilized for a portion of the simulations in this work.

Further studies of C5, C7, and C9 monolayers[7-10] involve investigation of the behavior of solid/liquid interface in a multilayer system, as well as a solid monolayer at coverages ranging from submonolayer through completion. In agreement with previous work, this work also indicates that C5 and C7 (as well as C9) crystallize in a solid RC phase at low temperatures. Unit cell parameters are proposed for submonolayer C5, C7, and C9 through X-ray diffraction patterns and a high coverage cell is proposed that involves molecules in fully commensurate positions on the graphite substrate that is determined by neutron diffraction of deuterated samples. Furthermore, one study from these authors also indicates "anomalous behavior[7]" in layers of C5, where the solid monolayer does not seem to coexist with a bulk fluid phase, similar to the other *n*-alkane systems studied, but melts very near to the bulk melting temperature of C5. For the purposes of this study, the cell parameters given in [10] that describe a "high coverage" unit cell are used in simulations to study the behavior of a highly commensurate monolayer (that is packed to supercede the intermolecular interaction to the substrate interaction) and compare that to a particular model proposed for 1.01 and 0.98 monolayers of C5 and C7 respectively.

To further underscore the importance and current interest in such systems, the only previous theoretical work that exists (to the authors knowledge) that studies the phase behavior or transitions in physisorbed pentane or heptane is a very recent MD study[11] that investigates the behavior of multilayer heptane in an (*NPT*) ensemble. Amongst other behavior reported in this work, the formation of gauche defects is monitored as a function of pressure near the solid-liquid phase transition, and is found to be a significant effect there. The formation of gauche defects as observed by these authors near the solid-liquid phase transition is relevant to this work for reasons that will be discussed in later sections.

Although theoretical studies of odd *n*-alkanes seems to be a topic of future interest, there is a wealth of work that exists over even-alkanes on graphite,[12-16] with a concentration focused on the study of hexane ($C_6H_{14}$) in particular. Most notably, the first study over hexane and butane proposes a "footprint reduction" mechanism[12] by which a phase transition is preempted by a space reduction in the monolayer that is a result of intermolecular scattering that supplies kinetic energy allowing the molecules to either tilt out of the surface plane, or else change conformation- in both cases decreasing their in-plane molecular footprint. This theory has been well adapted to these physisorbed systems, and most recent work over monolayer hexane on graphite[16] indicates that melting occurs primarily via the tilting mechanism, with only a very small contribution from gauche defects.

The purpose of this specific work is to study both systems of pentane and heptane monolayers, at two different coverages as to simulate two experimental determinations of a near-monolayer surface coverage, to understand better the phase transitions in these systems. Moreover, both experimental groups that have previously conducted studies over these systems have commented on the "sharpness"[6] or the "anomalous behavior"[7] that is observed in the solid-fluid phase transition in pentane monolayers on graphite as opposed to other physisorbed alkane monolayers (such as heptane). Thus, it is the purpose of this work to study how chain length in two physisorbed systems that are extremely similar in their solid phase behavior and have a very similar chain length, can exhibit such different behavior at the melting transition. Furthermore, this study will attempt to classify the phase transitions in terms of phase behavior and calculated thermodynamic quantities and distributions.

The reader should be informed at this point that understanding the phase transitions in monolayers of chain molecules is an arduous task, at best. Early theories of melting (such as the KTHNY theory[17]) are generally inapplicable to systems of chain molecules, as more degrees of freedom in the chains allow these molecules to exhibit three-dimensional motion, thus invalidating melting theories for simple 2D systems. Also, a direct study of the phase transition order is also somewhat ambiguous from simulations, as small system sizes can induce a variety of effects on energy fluctuations near the phase transitions which can be quite misleading[18]. To even be able to comment on phase transition orders in simulations, one has to consider a finite-size scaling scheme and study how the free energy scales with system size. Although the author acknowledges that this scheme would be extremely important to further our understanding of phase transitions in these monolayers, this work specifically is completed to study the melting behavior independent of order, consistent with the previous studies that have eluded to characterizing melting by the "footprint mechanism" which effectively characterizes melting behavior strictly in terms of molecular behavior, which is more applicable to simulations.

## II. SIMULATION DETAILS

A constant temperature, constant planar density, and constant molecule number ($N$, $\rho$, $T$) canonical ensemble molecular dynamics method is used to model pentane (C5), heptane (C7), and nonane (C9) monolayers physisorbed onto the basal plane of graphite. To model the C5, C7, and C9 molecules, the anisotropic united atom (AUA4)[19] model is used, which attempts to take into account the presence of hydrogen atoms by distinguishing the methyl ($CH_3$) groups from the methylene ($CH_2$) groups by shifting the pseudoatom centers more toward the hydrogen atoms. Unlike the united-atom (UA) model, which does not distinguish the two groups except for by mass, this model more accurately represents the intermolecular interactions present in the system, which is increasingly more important in smaller *n*-alkane molecules, where the ratio of methyl groups to methylene groups is comparable. Initially the UA model was used for this study, but a noticeable difference in behavior in the AUA model as compared to the UA model motivated an adaptation of the AUA4 model for this study. It is important to note that there is a significant difference in the behavior of these two models for odd alkane monolayers, which further contributes to recent speculation of such a difference[20] (since the AUA model gives the most "accurate" physical interpretation of an alkane).

The potential model utilized in this study is similar to that reported in previous work[16], therefore, only a topical description will be given here. In brief, both non-bonded and bonded interactions are considered. The bonded interactions are composed of three and four-body bending terms that describe motion about the bond-angles[21] and the dihedral angles[22] of the molecules. In addition, the non-bonded interactions are composed of (*i*) a simple 12-6 Lennard-Jones pair interaction potential (*ii*) Steele's potential[23], which describes the interaction between an atom and an infinite number of graphene layers. In general, the form of this potential model has been well-accepted to represent the phases and phase transitions of hexane monolayers and bilayers in previous studies.

To model the temperature dependence of the low-temperature solid phase, experimental determinations of unit cell parameters are used that correspond to *both* cell sizes and orientations given in [6] and [10]. From this data, the model chosen for both C5 and C7 monolayers is presented in table I, as well as the cell parameters utilized from [10] for C9 that involve a fully commensurate RC phase.

Matthies[6] proposes that one possible scenario for the solid crystalline phase observed involves a slight HB rotation, and after careful consideration of submonolayer cells proposed in [10] that are slightly extended in the *b*-axis direction of the unit cell (which could indicate a HB rotation), this work adopts a slight HB rotation for C5 and C7 monolayers that is reported in table I as well.

In all simulations, periodic boundary conditions are used, and temperature control is maintained by a velocity rescaling technique that rescales velocities such that equipartition is satisfied for the center-of-mass, rotational, and internal temperatures (for more detail, refer to [16]). The time step used in all simulations is 1 fs, and all simulations are typically carried out over a period of 700 ps, with the equilibration period spanning over the first 200 ps of the simulation. This equilibration period is found to be more than enough simulation time to achieve a thermodynamic equilibrium. To carry out the integration of the equations of motion, a velocity Verlet RATTLE[24] algorithm is utilized, which performs the integration while constraining the bond lengths to an equilibrium value of 1.535Å. The phase behavior is typically sampled in steps of 5K (in some cases 10K), and in steps of 2-3K near phase transitions, to delineate the temperature dependence of the phases and phase transitions.

## III. RESULTS

The results presented in this work reflect the main idea behind the different melting behavior observed for the C5 and C7 monolayers. For additional information on quantities that describe the different behavior, as well as visual aids, the reader is referred to [25]. The difference in melting behavior that is proposed in this work is emphasized in figure 1 through the structural order parameter, $OP_n$, defined as:

$$OP_n = \frac{1}{N_m} \left\langle \sum_{i=1}^{N_m} \cos n(\phi_i) \right\rangle. \tag{1}$$

This order parameter is presented for all coverages and temperatures of pentane and heptane monolayers studied. In eqn. (1), $\phi_i$ is the angle that the smallest moment of inertia axis of molecule $i$ makes with the $x$-axis of the computational cell. Due to the orientation of the molecular long axes along the $y$-direction of the cell, the values of $n$=2, 4, 6, and 8 are used to give information regarding the nature of the melting transition, where large values of $n$ are very sensitive to fluctuations and give information on the $n$-fold symmetry of the solid phase. In general, in the disordered (fluid) phase, the molecular orientations are randomly sampling angles in the substrate plane, so $OP_n$ vanishes in such a case. From figure 1, $OP_n$ seems to vanish very quickly near the melting temperature for both of the C5 monolayers, but seems to gradually decrease, beginning at a point significantly before the melting transition temperature, for the C7 monolayers. This indicates that, from a purely structural perspective, the C5 and C7 monolayers indicate differences in the temperature dependence of their orientational behavior with respect to the underlying substrate.

In figure 2, the center-of-mass bond orientational distribution, $P(a)$, is presented, which gives the probability of the center-of-mass of a neighboring molecule j, being a particular orientational angle $a$ (in degrees) from the center-of-mass position of a central molecule i. Pair distances are only taken over five consecutive neighbor shells, which are fixed through the atomic pair correlation function, $g(r)$. In a well ordered (solid) phase, the orientational correlation in figure 2 will have several distinct peaks corresponding to neighbors (within five neighbor shells) at fixed orientations from the central molecule. However, in a fluid phase, the correlation will become a straight line, as the center-of-mass positions will no longer be orientationally correlated. In figure 2, one can observe that the pentane monolayers go from a very well-ordered phase (with very well-defined peaks in $P(a)$) to a very disordered phase over a very narrow temperature region. However, the melting behavior of the C7 monolayer is a bit different, indicating a more gradual loss of order from the gradually decreasing peaks of $P(a)$ prior to, and after, the melting transition takes place. In addition, one can compare the atomic pair correlations (presented in [25]) for a complete picture of orientational and transitional order in the monolayer during the melting process for these two monolayers.

Figure 3 presents the three-dimensionally averaged static structure factor defined as:

$$S(q) = 1 + \int_0^\infty (g(r) - 1) \frac{\sin(qr)}{qr} r \, dr. \tag{2}$$

This function is obtained directly from the atomic pair correlation function, $g(r)$. This quantity is defined in figure 3 for monolayers of C5 and C7 that are simulated initially with a slight HB rotation, thus corresponding to data in [6]. In general, this quantity provides important information in reciprocal space that can be directly linked to the observed diffraction data by Matthies[6]. However, inspection of figure 3 would suggest that for the C7 and (mostly) C5 monolayers before melting, the atomic $g(r)$ has not yet completely converged, introducing the possible presence of truncation peaks in $S(q)$ at temperatures prior to melting. This can account for the three peaks in figure 3 before $q=1.5\text{Å}^{-1}$, which are not observed experimentally, even though the temperature dependence of the largest peak seems to be in good agreement with what experiment observes for both the C5 and C7 monolayers.

Additional data including specific heats, dihedral energies, snapshots of the melting process of C5, C7, and C9, as well as order parameters describing melting in C9 monolayers can be found in [25].

## IV. DISCUSSION

Observation of the data presented in figures 1-3 as well as that presented in [25] is evidence to the proposition that the melting behavior in the pentane and heptane monolayers studied in this work involves significant differences. This section will be devoted to (*i*) elucidating and discussing these differences for the pentane and heptane monolayers studied in this work, and (*ii*) proposing the origin for the observed difference, in comparison to STM data, studies of C9 monolayers, and studies of variations, most of which are presented in [25].

## A. Melting in Pentane and Heptane Monolayers

Although pentane and heptane are rather similar molecules, and the solid phase behavior of pentane and heptane monolayers are relatively similar, the melting behavior of these monolayers is observed to be different in experiment, as well as in simulations. Figure 1 shows the order parameter, $OP_n$, for both pentane and heptane monolayers studied in this work. Interestingly, it is evident that both pentane monolayers indicate a very rapid transition over a narrow temperature region, and both heptane monolayers exhibit a more gradual phase transition over a more broad temperature region. Comparing this observation to the bond-orientational correlations presented in figure 2, one observes that this differing loss of orientational order of these two monolayers takes place (*i*) with respect to a fixed axis position (as is evidenced in figure 1), and (*ii*) with respect to neighboring molecular center-of-masses (evidenced from figure 2). These two figures suggest the notion that the phase transitions in the pentane and heptane monolayers seem to be significantly different, based upon the loss of order with respect to the graphite substrate as the temperature is increased. In addition, table II presents melting temperatures of the studied monolayers (including nonane) in comparison to what experiment suggests. In most cases, the melting temperatures predicted by simulations are in very good agreement with experiment, which means that simulations suggest the same nature of the melting behavior as experiment has previously observed (i.e. the "sharp" or anomalous" behavior of C5 monolayers, and the more gradual nature of the C7 monolayers), and simulations suggest similar melting transition temperatures as experiment has observed.

Also, from figure 3, and calculated from the atomic pair correlations [25], the three-dimensionally averaged static structure factor is shown for C5 and C7 monolayers that are simulated in connection to work done by Matthies[6]. The temperature dependence of the static structure factor for the C7 monolayer indicates a peak at ca. $q$=1.45Å$^{-1}$ which broadens at about 180K in excellent agreement with the temperature dependence shown in experiment, where there is a slight broadening of the peak intensity between $T$=175K and $T$=180K. Also, the slight shift in this peak toward higher $q$ as the temperature is increased toward melting is also in very good agreement. Further comparison of the C5 monolayer indicates less fair agreement in this aspect, although good qualitative agreement. First of all, due to the diverging nature of the integral performed to obtain $S(q)$ in this study, peaks at low-$q$ (below what is shown) can not be readily compared to experiment. In this case, the peak at higher $q$ (corresponding to $q$ slightly higher than 1.4Å$^{-1}$) in figure 5 occurs about $q$=1.5Å$^{-1}$, but still indicates the same qualitative behavior of the temperature dependence given by Matthies of the peak labeled (11), with a very sharp decrease of intensity at the melting transition (which occurs between 99-105K in experiment, compared to 92K in simulations), and a pretty consistent peak positioning through all temperatures. Also, simulations indicate that there are three peaks below this (11) peak, which can be attributed to artificialities present due to the truncation of $g(r)$ at 30Å, where the pair correlation function has not yet quite converged. However, the rapid disappearance of the distinct nature of all peaks at temperatures between 90-95K is in agreement with experimental observations of such behavior between temperatures of 99-105K[6].

## B. Gauche Defects and Nonane Monolayers

So far, the emerging picture of the melting behavior in the C5 and C7 monolayers is one that differs in both experiment and simulations. However, it is at this point where one may consider why this melting behavior is different. Since the monolayers studied in this work are nearly identical in every respect except for the length of the molecules making up the monolayer, one may initially consider the role that the alkane chain length plays in this disordering process. In particular, as the alkane chain length becomes longer, the affinity for the molecule to assume a gauche conformation becomes greater, as there are 2($n$-3) possible gauche defects allowed to a molecule, where $n$ is the total number of carbon atoms in the alkane chain. In addition, snapshots of the C7 monolayer[25] show that there are more molecules with gauche defects in the alkyl chains as the temperature is increased toward and above the melting transition temperature. Therefore, the first step taken to understand this difference in the melting transition was to completely eliminate gauche defects in the alkyl chains. One would expect that if gauche defects were responsible for the gradual nature of the melting transition in the C7 monolayer as compared to the C5 monolayer, and the C5 monolayer did not depend on gauche defects (see figures 6 and 7 in [25] for more info here), then elimination of the formation of gauche defects in the C7 monolayer would yield a very sharp phase transition over a narrow temperature region which is the observed result for C5 monolayers. Thus, to eliminate gauche defects, the constants in the series describing the dihedral potential are multiplied by 10 in order to make it energetically unfavorable for gauche defects to form in the alkyl chains. Curiously, the phase transition for the C7 monolayers without gauche defects is still gradual[25] as evidenced by order parameters similar to that presented in figure 1. However, it is interesting to note that the melting transition temperature of the C7 monolayer without gauche defects was increased by ca. 60K with the elimination of gauche defects, indicating a stronger dependence of the melting transition on gauche defects than previously studied hexane monolayers[16], where only a 20K increase in the melting temperature was observed in such a case. In fact, it is intuitive that gauche defects be responsible for this gradual behavior, since the heptane monolayers are inherently more dependent on gauche defects than the pentane monolayers, and the formation of gauche defects with increasing temperature would lead one to intuitively suggest a more gradual transition (if the transition took place from vacancy formation based upon gauche defects, such as the footprint reduction theory[12] proposes). However, simulations tend to suggest that such is *not* the case.

The next step taken to understand this gradual melting behavior was to study the melting transition in the C9 monolayer, and compare the melting behavior of this monolayer with that already discussed for the C5 and C7 monolayers. Interestingly enough, the melting transition in this monolayer exhibits the same type of gradual melting behavior as that of the C7 monolayer, which once again indicates the role of chain length in the monolayer on the melting transition. However, comparing snapshots of the C5, C7, and C9 monolayers (presented in [25]) indicates quite different molecular behavior near the melting transition. In the case of C5 monolayers, about 5K below the melting transition, the molecules seem to be primarily in a well-defined solid (with some molecules slightly rotated). However, in the C7 and C9 monolayers, at temperatures well below the melting transition (approaching 40K!), it is evident that the molecules start to gradually "shift" by moving about their molecular long axes in the lines of lamellae that they are in. As the temperature is increased, the shifting becomes more apparent, and it seems that at temperatures near the melting transition, the shifting of the lamellae is responsible for disordering the monolayer, even though there still seems to be some order in the monolayer after the transition due to melting in this way (which is evident from figure 2).

Such an observation of the melting behavior brings forth an important point about this system, which is that, as a direct result of chain length, monolayers with molecules of length C7 and greater involve a disordering process that depends primarily upon the effect of "sliding" or translational motion about the molecular long axis of the molecule. In fact, such an observation is not unreasonable given a recent STM study of tricosane (C23) on graphite[26] where the authors of this study even comment on the "shift of the molecules" in the solid phase. This would then suggest that the difference in the melting behavior in the C5 and C7 monolayers is due to the difference in the motions that are responsible (and energetically favorable), and contribute to the eventual disordering of the monolayer. For the shorter alkane (C5), the most energetically favorable way for the molecule to disorder with respect to the graphite substrate would be orientational motion about the molecular center of mass, resulting in a loss commensurability with the substrate. However, the author speculates that the difference in chain length between a linear molecule with 5 carbon atoms and 7 carbon atoms in the chain seems to be enough to stifle the motions associated with the "sharp" phase transition, and require that the molecule disorder by translational motion of the molecule about its molecular long-axis. Such a difference in the motions associated with the phase transition seems to have a large role in the different melting in these monolayers that have been studied in this work.

Finally, this type of translational motion prior to melting in the C7 and C9 monolayers could relate to observations of alkanes on a variety of different surfaces. Besides this type of "sliding" effect being observed in STM images of C23[26] on graphite and C25 and C50[27] on graphite in previous work, alkanes with lengths as large as C60 on graphite[28], and those such as C44 on Cu(100) and Cu(110)[29] have been observed to exhibit solid phase behavior consisting of lamellae, in some sense similar to that of C5, C7, and C9. Hentschke et al[27] propose that the solid phase in longer alkanes on graphite involves "rod-like" molecules on the surface, without the "coiled" molecular behavior that is observed in the bulk fluid. In fact, simulations in this particular study seem to suggest similar behavior, and due to the strict solid RC phase, the C7 and C9 monolayers are stifled from orientational disorder, and exhibit "sliding" motion about their long axes. In this sense, this type of behavior exhibited for C7 and C9 seems to be in registry with what has been observed in longer alkanes in previous experiment, and the relation between an increasing chain length, the consistent intermolecular spacing in the solid RC phase, and the means by which the monolayer disorders could be useful to understand the molecular and orientational behavior associated with the disordering of alkane monolayers, among a variety of both alkane chain lengths and substrates.

# V. CONCLUSIONS

The study presented in the previous sections brings some very key conclusions to the behavior of C5 and C7 monolayers near the melting transition, in correspondence with two previous experimental studies. In particular, this study proposes the following conclusions: (*i*) The melting behavior in physisorbed monolayers of C5 and C7 on graphite indicate significant differences. In particular, this study finds that the melting transition for C5 monolayers involves a very rapid loss of order with respect to the substrate, whereas the melting transition for C7 monolayers seems to be one of a very gradual nature. This is in good agreement with the "anomalous behavior"[7] and the "sharpness"[6] observed experimentally for C5 monolayers. (*ii*) On the basis of comparison of the C5 and C7 monolayers, the differences between C5 and C7 (namely, the difference in torsion and chain length) are varied to study how each contributes to the "gradual" behavior of the melting transition. This study proposes that an increased chain length in the solid RC phase "stifles" the orientational disorder of the longer molecule, which contributes to this gradual melting behavior for the longer adsorbed molecules due to more uniaxial sliding of sublattices and molecules within sublattices. (*iii*). This study proposes melting behavior (with respect to melting temperatures) that seems to be in good agreement with experiment. This is the first study conducted regarding the solid-state behavior of physisorbed alkanes with the AUA4 model, and thus validates its use for further studies over these types of systems. (*iv*) Finally, the translational motion prior to melting exhibited by C7 and C9 monolayers is in agreement with that presented in previous experimental studies, indicating that this melting behavior could be inherent to many more physisorbed alkanes on surfaces (whose phase behavior consists of ordered lamellae). Furthermore, whereas previous understanding of melting comes from a "space reduction,"[12] (which this study predicts to be valid for odd-alkanes as well) it is possibly more informative to understand this "sliding" type of behavior at high temperatures prior to melting, as the author speculates that it is *because* of this sliding that gauche defects form in the ends of the physisorbed molecules, providing the space reduction. Therefore, this uniaxial sliding behavior could present a more fundamental understanding of the melting behavior in these monolayers than "space reduction," which could ultimately be a consequence of the observed translational motion.

## ACKNOWLEDGEMENTS

The author is grateful to Paul Gray and the UNI CNS for use of computing facilities. Also, the author acknowledges the UNI Physics Dept. for use of computing facilities and John Deisz for helpfulness with operation of these facilities.

# TABLES AND FIGURES

**Table I.** Simulated unit-cell parameters, computational cell sizes, and numbers of molecules and atoms in simulations for pentane, heptane, and nonane monolayers. Note that a (*) refers to a fully commensurate monolayer, in which cell parameters are taken from experimental work[10]. Other cell parameters are those proposed by Matthies[6]. Initial HB rotations for those monolayers given by Matthies is ±3° and ±6° for the C5 and C7 monolayers respectively.

|          | $a$ (Å) | $b$ (Å) | cell size (Å)   | $N_m$ | $N_a$ |
|----------|---------|---------|-----------------|-------|-------|
| Pentane  | 16.98   | 4.66    | 67.92 X 74.56   | 128   | 640   |
| Pentane* | 17.1    | 4.26    | 68.88 X 68.3    | 128   | 640   |
| Heptane  | 21.9    | 4.57    | 65.7 X 73.12    | 96    | 672   |
| Heptane* | 22.0    | 4.26    | 66.0 X 76.68    | 108   | 756   |
| Nonane*  | 27.0    | 4.26    | 81.0 X 68.16    | 96    | 864   |

**Table II.** Melting transition temperatures ($T_m$) with respective uncertainty for each monolayer studied (including nonane) compared to experimental observations of melting temperatures[6,10] in each monolayer (if this is studied). Fully commensurate monolayers (with unit cell parameters) are indicated with an asterisk (*).

| alkane monolayer | Simulated $T_m$ | Experimental $T_m$ |
|---|---|---|
| Pentane | 92 ± 3K | 99-105K[6] |
| Pentane* | 147 ± 3K | 150K[7] |
| Heptane | 178 ± 5K | 170 ± 10K[6] |
| Heptane* | 218 ± 5K | 206K[8] |
| Nonane* | 255 ± 3K | N/A |

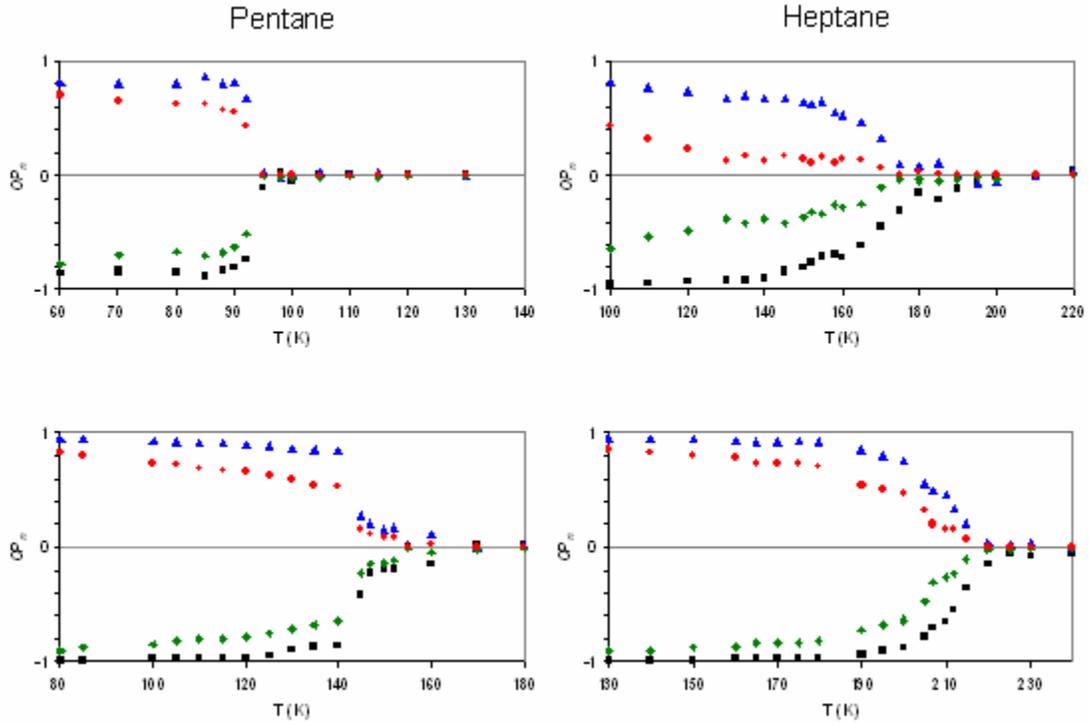

(color online) **Figure 1.** *OP_n*, with *n*=2, 4, 6, and 8, for pentane monolayers (left two panels) and heptane monolayers (right two panels). In each panel, the corresponding symbol is: for *n*=2 (black squares), *n*=4 (blue triangles), *n*=6 (green diamonds), and *n*=8 (red circles). The monolayers with cell parameters given by Matthies[6] are on the top two panels, and those with cell parameters given by Arnold et al.[10] (for a fully commensurate monolayer) are in the bottom two panels.

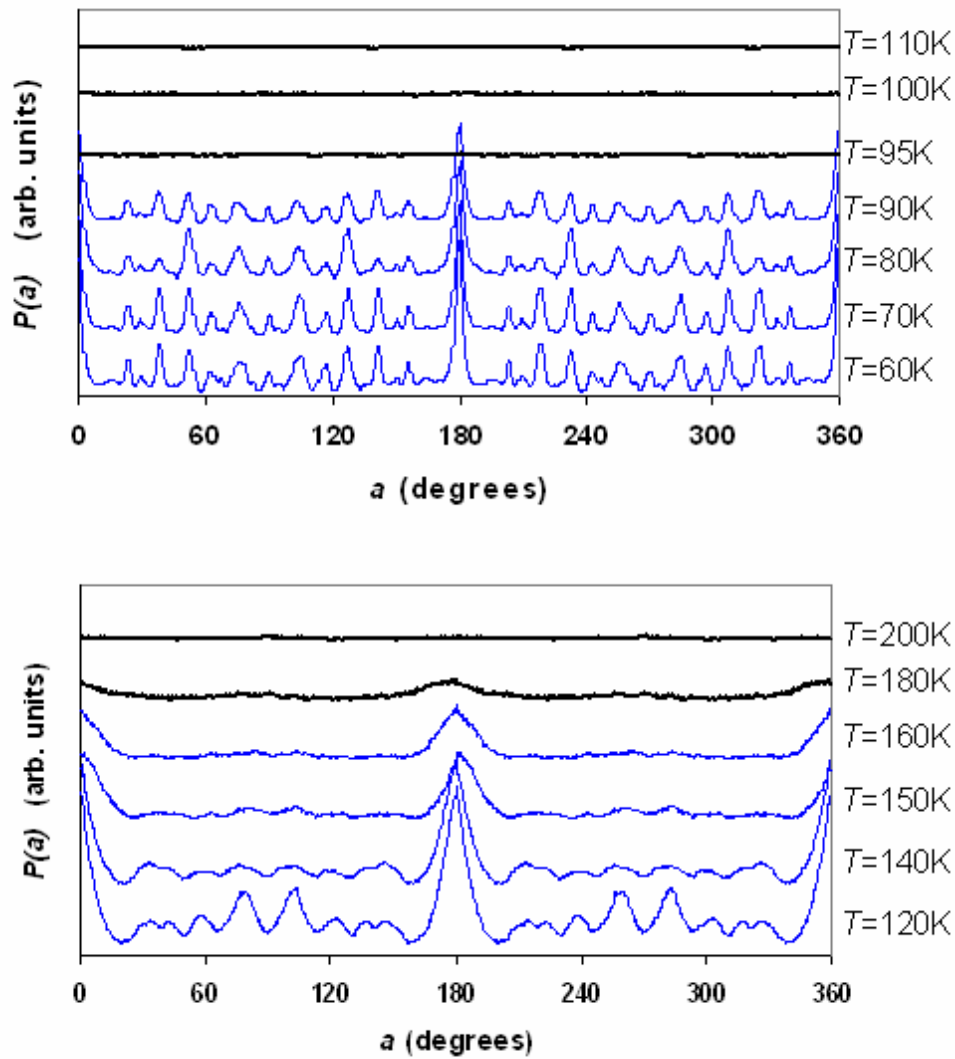

(color online) **Figure 2.** Bond-orientational distributions, $P(a)$, for $1^{st}$-$5^{th}$ neighbor shells for monolayers of pentane (top panel) and heptane (bottom panel) with unit cell parameters given by Matthies[6]. Note the very sudden loss of long-range orientational order between 90-95K in the pentane monolayer, and the more gradual loss of long-range orientational order in the heptane monolayer.

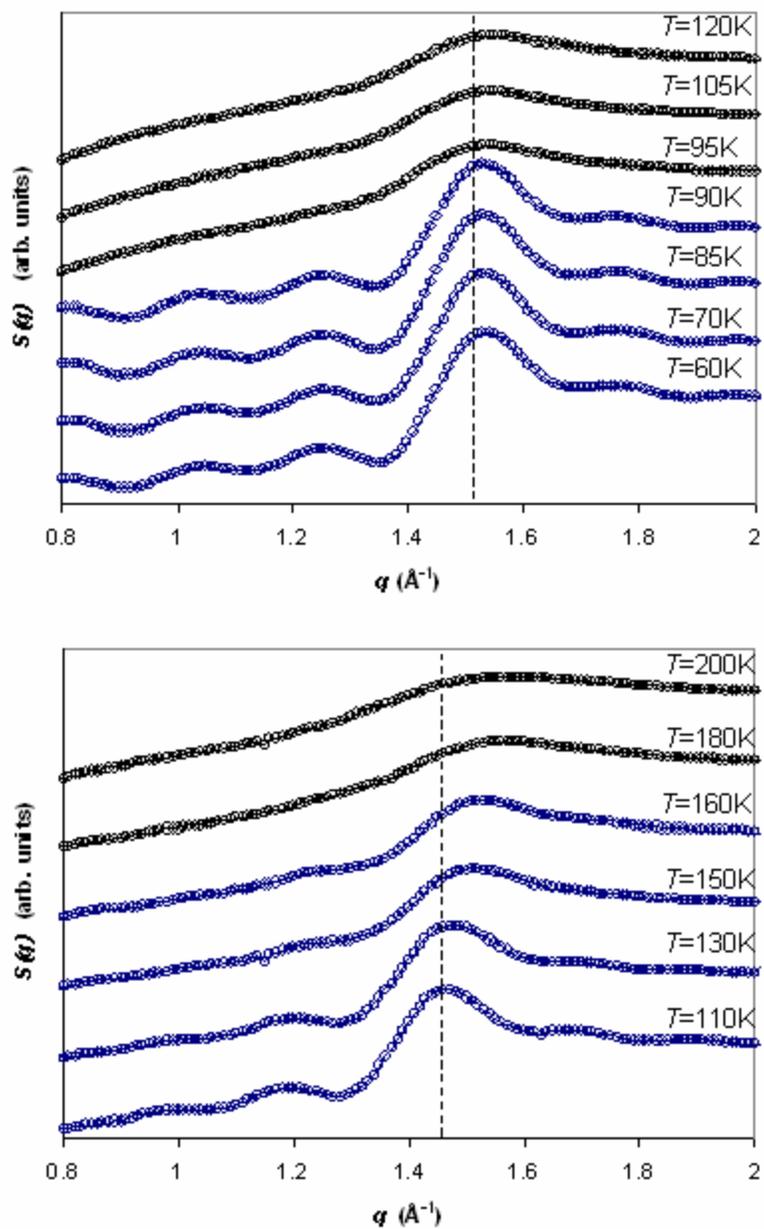

(color online) **Figure 3.** Static structure factors, $S(q)$, for pentane (top panel) and heptane (bottom panel) monolayers simulated from experimental observation by Matthies[6]. Note the slight shift to higher $q$ observed for the middle peak in heptane monolayers, and the relatively strict positioning of the peak in pentane monolayers, even after melting.